\documentclass{ws-ijmpd}
\usepackage[super,compress]{cite}
\usepackage{graphicx}
\usepackage{hyperref}
\usepackage{subeqnarray}
\usepackage{color}
\usepackage[normalem]{ulem}
\usepackage{url}
\usepackage{multirow}
\usepackage{array}
\newcommand{\beq}{\begin{equation}}
\newcommand{\eeq}{\end{equation}}
\newcommand{\beqn}{\begin{eqnarray}}
\newcommand{\eeqn}{\end{eqnarray}}
\newcommand{\beqs}{\begin{subeqnarray}}
\newcommand{\eeqs}{\end{subeqnarray}}
\newcommand{\nn}{\nonumber}

\begin{document}

\markboth{Oliveira, Benone, Almeida, Crispino}
{Analytical investigation of wave absorption by a rotating black hole analogue}

%
\catchline{}{}{}{}{}
%

\title{ANALYTICAL INVESTIGATION OF WAVE ABSORPTION BY A ROTATING BLACK HOLE ANALOGUE
}

\author{LEANDRO A. OLIVEIRA$^\ast$ AND CAROLINA L. BENONE$^\dagger$
}

\address{Campus Salin\'opolis, Universidade Federal do Par\'a,\\
68721-000, Salin\'opolis, Par\'a, Brazil\\
$^\ast$laoliveira@ufpa.br\\
$^\dagger$benone@ufpa.br}

\author{AMANDA L. ALMEIDA$^\ddagger$ AND LUÍS C. B. CRISPINO$^\S$}

\address{Faculdade de F\'{\i}sica, Universidade Federal do Par\'a\\
66075-110, Bel\'em, Par\'a, Brazil\\
$^\ddagger$amanda.almeida@icen.ufpa.br\\
$^\S$crispino@ufpa.br}

\maketitle

\begin{history}
\received{Day Month Year}
\revised{Day Month Year}
\end{history}

\begin{abstract}
Perturbations in a draining vortex can be described analytically in terms of confluent Heun functions. In the context of analogue models of gravity in ideal fluids, we investigate analytically the absorption length of waves in a draining bathtub, a rotating black hole analogue, using confluent Heun functions. We compare our analytical results with the corresponding numerical ones, obtaining excellent agreement. 
\end{abstract}


\ccode{PACS numbers: 04.70.-s, 04.30.Nk, 43.20.+g, 47.35.Rs}


\section{Introduction}

Black holes (BHs) are extreme objects, characterized by the presence of an event horizon~\cite{wald_1984}. BHs of the Kerr family have a spacetime singularity in their core, and singularity theorems predict that they can be formed by the collapse of physical matter \cite{Penrose:1964wq}. 
Event horizons and singularities are believed to be linked to quantum physics, what makes BHs keystones in the development of a complete quantum description of the gravitational interaction. 

By studying the quantum nature of BHs, Hawking has shown that these objects emit radiation as black bodies \cite{Hawking:1974rv}, the so-called \textit{Hawking radiation}. 
The temperature associated with the Hawking radiation is very feeble, being of the order of $10^{-6}(M_\odot/M)$K, where $M$ is the mass of the BH and $M_\odot$ is the mass of the Sun. This result, together with the nature of the event horizon, turns the task of measuring Hawking radiation directly into an extremely difficult one. In order to circumvent such difficulty, Unruh proposed that, instead of working directly with BHs, one could investigate systems that have the same kinematic properties of a BH, but which are easier to access~\cite{Unruh:1980cg}. 

With his pioneering work, Unruh gave venue to a new field of investigation: the analogue gravity. The first analogue model proposed by Unruh consisted of a fluid with a sink in the middle, such that acoustic perturbations in this fluid ``feel'' an effective curved geometry \cite{Unruh:1980cg}. In this case, the event horizon is formed when the velocity of the fluid equals the speed of sound. This system was generalized to include rotation~\cite{Visser:1997ux}, resulting in an effective BH spacetime that presents an ergorregion, as the Kerr BH. 

Although these acoustic systems are among the most widely known analogue models, they present a strong disadvantage: In order for the event horizon to form, the fluid has to be supersonic in a certain region. This fact makes such a system difficult to achieve without creating turbulence in the fluid.

A different analogue BH model was proposed by Sch\"{u}tz\-hold and Unruh \cite{Schutzhold:2002rf}, composed by gravity waves in a shallow basin. In this case, the velocity of propagation of the waves depends on the depth of the basin, such that the speeds involved are of the order of $m/s$. 

Along the years, several other analogue models of gravity have been proposed (cf., e. g.~\citen{Barcelo:2005fc,Garay:1999sk,Hod:2014hda,Oliveira:2014oja,Benone:2014nla,Oliveira:2015vqa,Cardoso:2016,Benone:2018}).
Recently, claims of experimental verification of Hawking radiation in analogue systems have been made~\cite{Weinfurtner:2010nu, Steinhauer:2015saa}. Although there is some dispute whether these observations were in fact of Hawking radiation \cite{Dardashti:2015tgp}, such developments give additional reasons to study BH physics in analogue gravity. 

It has been shown that Hawking radiation is closely related to BH absorption~\cite{Hawking:1974sw}, a subject that has been extensively studied in the literature (cf., e. g.~\citen{ACS:2005, ACS:2010, ACS:2011, Benone:2014qaa+Ad, Benone:2016, ACS:2017}). 
Although the equations related to wave propagation in fluids can be somewhat simplified, the solutions for the full range of frequencies have mainly been treated numerically (cf., e. g.,~\citen{Oliveira:2010zzb,Dolan:2011zza}), with few exceptions published in the literature (cf., e. g.,~\citen{Vieira:2014rva}).

We study fluid perturbations in a draining vortex, finding the absorption length analytically in terms of confluent Heun functions. We also compare our analytical results with numerical ones. The remainder of this letter is organized as follows. In Sec. \ref{sec:spacetimes} we review the main aspects of the effective spacetime of the draining bathtub. In Sec. \ref{sec:perturbations} we consider linear perturbations in the draining bathtub. In Sec. \ref{sec:absorption} we obtain absorption length. In Sec. \ref{sec:analytical} we present analytical results for the wave absorption. In Sec. \ref{sec:con} we consider the low-frequency limit and the numerical solution for the absorption length, as consistency checks of our analytical solution. We present our conclusion in Sec. \ref{sec:conclusion}.

\section{Effective spacetime of a draining bathtub}
\label{sec:spacetimes}
In fluid mechanics, we can use mass conservation and Newton's second law to describe an inviscid fluid flow by the continuity equation 
\beq
\frac{\partial \rho}{\partial t}+\nabla \cdot\left(\rho \, \vec{v}\right)=0,
\label{eq:cont}
\eeq
and Euler equation
\beq
\frac{\partial \vec{v}}{\partial t} +\left(\vec{v} \cdot \nabla\right)\,\vec{v} +\frac{\nabla P}{\rho}=0,
\label{eq:euler}
\eeq
where $\rho$ is the mass density, $P$ is the pressure and $\vec{v}$ is the flow velocity of the fluid.
A barotropic fluid flow obeys the following equation of state:
\beq
P=P\left(\rho \right).
\label{eq:baro}
\eeq

For an irrotational flow, the velocity $\vec{v}$ satisfies
\beq
\nabla \times \vec{v}=0,
\label{eq:irrot}
\eeq
so that it can be represented as a conservative field, namely
\beq
\vec{v}=-\nabla \Phi,
\label{eq:conser}
\eeq
where $\Phi$ is the (flow velocity) potential of the fluid.

The flow velocity of the vortex with a sink can be written as
\beq
\vec{v}=v_r(r) \,\hat{r}+v_\theta(r) \,\hat{\theta}.
\label{eq:flow_velo_1}
\eeq
We assume a steady non-perturbed fluid flow (without an explicit time dependence) and that the flow velocity components are functions of the radial coordinate only.

Substituting the flow velocity of the vortex, given by Eq.~\eqref{eq:flow_velo_1}, into Eqs.~\eqref{eq:cont} and~\eqref{eq:irrot}, we find that
\beq 
\vec{v}=-\frac{D}{r} \hat{r}+\frac{C}{r} \hat{\theta},
\label{eq:flow_velo_2}
\eeq
where $D$ and $C$ are positive constants related, respectively, with the draining of the fluid to the center of the vortex (sink) and with its circulation (the circulation is $\Gamma = 2\,\pi\,C$).

Linear perturbations in the fluid flow described by Eqs. \eqref{eq:cont}, \eqref{eq:euler}, \eqref{eq:irrot} and \eqref{eq:baro} can be represented as a Klein-Gordon equation~\cite{Unruh:1980cg}, namely 
\beq
\nabla_\mu \nabla^\mu \phi = \frac{1}{\sqrt{|g|}} \partial_\mu \left( \sqrt{|g|} g^{\mu \nu} \partial_\nu \phi \right) = 0, \label{eq:Klein}
\eeq
in an effective curved geometry with contravariant metric components $g^{\mu\nu}$ and determinant $g$.

The scalar function $\phi$ in Eq. (\ref{eq:Klein}) is related to the perturbation of the flow velocity $\delta\vec{v}$ by 
\beq
\delta\vec{v}= -\nabla \phi, 
\label{eq:pert}
\eeq
and the speed of the perturbation can be written as
\beq
c\equiv\sqrt{dP/d\rho}.
\label{eq:speed}
\eeq
From now on we set $c\equiv 1$.

The line element of the effective spacetime of the draining bathtub may be written as~\cite{Berti:2004ju}
\beq
ds^2=-f(r)dt^2+f(r)^{-1}dr^2+\left(rd\theta -\frac{C}{r} dt \right)^2,
\label{eq:elem_1}
\eeq
with 
\beq
f(r) \equiv 1-\frac{D^2}{r^2}.
\label{eq:lapse}
\eeq
The event horizon is located where the radial flow velocity equals the speed of the perturbations, i.e. at 
\beq
r_{\rm h}=|D|. 
\eeq
The ergoregion is defined as the region where the modulus of the flow velocity is larger than the speed of the perturbation. The outer boundary of the ergoregion is given by 
\beq
r_{\rm e}=\sqrt{D^2+C^2}.
\eeq

\section{Linear perturbations and their analytical description}
\label{sec:perturbations}
We may describe the linear perturbations in the draining bathtub using the following {\it ansatz}
\beq
\phi(t,\vec{r})=\frac{1}{\sqrt{r}}\sum_{m=-\infty}^{\infty}\psi_m(t,r)\exp\left(im\,\theta\right),
\label{eq:ansatz}
\eeq
where $m$ is an integer, the azimuthal number, associated with the angular momentum of the perturbation.

Substituting Eq.~\eqref{eq:ansatz} into Eq.~\eqref{eq:Klein}, we find the following partial differential equation (PDE)
\beq
\left[ -\left(\frac{\partial}{\partial t} + \frac{i C m}{r^2} \right)^2 + f\frac{\partial}{\partial r}\left(f\frac{\partial}{\partial r}\right) - V_m(r) \right] \psi_m(t,r) = 0 ,  
\label{eq:pde}
\eeq
that describes the perturbation in $(t, r)$ coordinates.

The function $V_m(r)$ is the effective potential that the waves are submitted, given by
\beq
V_m(r) \equiv f(r) \left( \frac{m^2 - 1/4}{r^2} + \frac{5D^2}{4r^4} \right).
\label{eq:potential}
\eeq
From Eqs.~\eqref{eq:lapse} and~\eqref{eq:potential}, it can be found that the effective potential goes to zero at the event horizon and at infinity.

The PDE given by Eq.~\eqref{eq:pde} describes the perturbation in the time-domain of the effective spacetime. 
We constrain the dependence of the perturbation on its oscillation frequency $\omega$, 
as follows~\cite{Dolan:2010zza,Dolan:2011ti} 
\beq
\psi_m(t,r)=u_{\omega m}(r)\exp\left(-i\omega t\right). 
\label{eq:tfunc}
\eeq
From Eqs. \eqref{eq:pde} and \eqref{eq:tfunc}, we may obtain an ordinary differential equation (ODE), namely
\beq
\left[ f\frac{d}{d r}\left(f\frac{d}{dr}\right) + \left(\omega - \frac{mC }{r^2} \right)^2 - V_m(r) \right] u_{\omega m}(r) = 0.
\label{eq:radeq}
\eeq
Note that Eq.~\eqref{eq:radeq} involves first- and second-order derivative terms. In order to eliminate the first derivative term, we may define the Regge-Wheeler coordinate $x$ as follows
\beq
\frac{d}{dx} \equiv f\frac{d}{dr},
\eeq
or, explicitly,
\beq
x=r+\frac{r_{\rm h}}{2}\ln\left|\frac{r-r_{\rm h}}{r+r_{\rm h}}\right|.
\label{eq:regge}
\eeq
From Eq.~\eqref{eq:regge}, we notice that at the event horizon the Regge-Wheeler coordinate goes to negative infinity. Furthermore, as the $r$-coordinate goes to infinity, the Regge-Whee\-ler coordinate also goes to infinity.

Using the Regge-Wheeler coordinate, Eq.~\eqref{eq:radeq} can be rewritten as follows:
\beq
\left[ \frac{d^2}{dx^2} + \left(\omega - \frac{C m}{r^2} \right)^2 - V_m(r) \right] u_{\omega m}(x) = 0.
\label{eq:radeq1}
\eeq

Since the ODE~\eqref{eq:radeq1} does not depend on the first-order derivative related to $x$-coor\-di\-na\-te, it is possible to show that the Wronskian of two linear independent solutions of this ODE does not depend explicitly of the  $x$-coordinate. Thus, we may write the following equation for the Wronskian $W[u(x),\bar{u}(x)]$ of the solutions at event horizon and spatial infinity of Eq.~\eqref{eq:radeq1}, namely
\beq
W[u(x),\bar{u}(x)]\Big|_{x\rightarrow -\infty}=W[u(x),\bar{u}(x)]\Big|_{x\rightarrow \infty},
\label{eq:wrons}
\eeq
where
\beq
W[g(x),h(x)] \equiv g(x)\frac{d}{dx}h(x)-h(x)\frac{d}{dx}g(x)
\eeq
and $\bar{u}(x)$ is the complex conjugate of $u(x)$.

\section{Absorption cross section}
\label{sec:absorption}

In this Section we describe the absorption of waves by the draining vortex. In order to describe the absorption process, the following boundary condition at spatial infinity (very far from the event horizon) holds 
\beq
u_{\omega m}(x)\Big|_{x\rightarrow \infty}=A^{\rm in}_{\omega m}\exp(-i\omega x)+A^{\rm out}_{\omega m}\exp(i\omega x),
\label{eq:inf}
\eeq
where the coefficients $A^{\rm in}_{\omega m}$ and $A^{\rm out}_{\omega m}$, respectively, are associated with the ingoing and outgoing waves. At the horizon, we have
\beq
u_{\omega m}\left(r\right)\Big|_{r\rightarrow r_{\rm h}}=\exp[-i\widetilde{\omega}x], 
\label{eq:hor}
\eeq
where
\beq
\widetilde{\omega}\equiv\omega-\frac{mC}{r_{\rm h}^2}. 
\label{wt}
\eeq

We may combine the coefficients $A^{\rm in}_{\omega m}$ and $A^{\rm out}_{\omega m}$ to write the transmission and the reflection coefficients, respectively, as
\beq
|{\mathcal T}_{\omega m}|^2\equiv \frac{1}{|A^{\text{in}}_{\omega m}|^2}
\label{eq:trans}
\eeq
and
\beq
|{\mathcal R}_{\omega m}|^2\equiv \left|\frac{A^{\text{out}}_{\omega m}}{A^{\text{in}}_{\omega m}}\right|^2,
\label{eq:refle}
\eeq
which satisfy the following relation
\beq
|{\mathcal R}_{\omega m}|^2+\frac{\widetilde{\omega}}{\omega}|{\mathcal T}_{\omega m}|^2=1.
\label{eq:tr}
\eeq
Using Eqs.~\eqref{eq:wrons},~\eqref{eq:inf},~\eqref{eq:hor} and~\eqref{eq:refle}, we may obtain the following expression
\beq
|{\mathcal R}_{\omega m}|^2=1-\frac{4}{{\mathcal S}_{\omega m}+2},
\label{eq:res}
\eeq
with
\beq
{\mathcal S}_{\omega m} \equiv \frac{\omega}{\widetilde{\omega}}\left[\left| u_{\omega m}\left(r\right)\right|^2+\frac{1}{\omega^2}\left| \frac{du_{\omega m}}{dr}\right|^2\right]_{r\rightarrow \infty}.
\eeq

The total absorption length is given by
\beq
\sigma_{\rm abs}=\sum_{m=-\infty}^{\infty}\sigma^{(m)}_{\rm abs},
\label{eq:totab}
\eeq
where $\sigma^{(m)}_{\rm abs}$ is the partial absorption length, obtained for each value of azimuthal number $m$, which can be written in terms of the reflection coefficient as
\beq
\sigma^{(m)}_{\rm abs}=\frac{1}{\omega}\left(1-|{\mathcal R}_{\omega m}|^2\right).
\label{eq:mab}
\eeq

We may express the absorption length in terms of ${\mathcal S}_{\omega m}$ as
\beq
\sigma_{\rm abs} = \frac{4}{\omega}\sum_{m=-\infty}^{\infty} \frac{1}{{\mathcal S}_{\omega m}+2}.
\label{eq:ab}
\eeq

\section{Analytical results}
\label{sec:analytical}
The analytic solutions of Eq.~(\ref{eq:radeq}) can be expressed in terms of confluent Heun functions, which are solutions of confluent Heun equations~\cite{Ronveaux:1995}. We may write
\beq
u_{\omega m}(r)=r^{\frac{5}{2}}\left(K_1\,\chi^+ +K_2\,\chi^-\right),
\label{eq:wavefunction}
\eeq
where
\beq
\chi^\pm \equiv (D^2 z)^{\pm\frac{\beta}{2}}\,{\rm HeunC}\left(\alpha,\, \pm \beta,\, \gamma,\, \delta,\, \eta; \,z\right),
\eeq
with 
\beqn
&&z \equiv 1-r^2/r_{\rm h}^2,\\
&&\alpha=0,\\
&&\beta\equiv i\,r_{\rm h}\,\widetilde{\omega},\\
&&\gamma=1,\\
&&\delta \equiv -\frac{1}{4}r_{\rm h}^2\,\omega^2,\\
&&\eta \equiv\frac{1}{4}\left[r_{\rm h}^2\,\omega^2-m^2\,\left(\frac{C^2}{r_{\rm h}^2}+1\right)+2\right].
\eeqn
The constants $K_1$ and $K_2$ are obtained considering the boundary condition~(\ref{eq:hor}) at the event horizon and its derivative.

The confluent Heun functions present two regular singular points, at $z=0$ and $z=1$, and one irregular singular point, at $z=\infty$. The confluent Heun equation arises when we perform a confluence process in the general Heun equation, what decreases the number of parameters.

\section{Consistency Checks}
\label{sec:con}
In this section we check our analytical results in two different ways: (i) Considering a low-frequency approximation and (ii) comparing with the numerical results.

\subsection{Low-frequency approximation}
To obtain an analytic expression for the absorption length in the low-frequency approximation, we take the limit $\omega \rightarrow 0$ in Eq.~\eqref{eq:wavefunction}, obtaining
\beqn
\lefteqn{{\rm HeunC}\left(0,\, \mp\varpi,\, 1,\, 0,\, \frac{1}{4}\left[2-m^2\left(\frac{C^2}{r_{\rm h}^2}+1\right)\right]; \,z\right) =} \nn\\ 
&&\left(\frac{1}{1-z}\right)^{1\pm \lambda} {\rm F}\left(1\pm\lambda, \pm \lambda; 1\mp \varpi; \frac{z}{z-1}\right),
\label{eq:heunlf}
\eeqn
where 
\beq
\lambda \equiv \frac{m}{2}\left(1-\frac{iC}{r_{\rm h}}\right), \hspace{1cm} \varpi \equiv \frac{imC}{r_{\rm h}},
\eeq
and ${\rm F}(b,c;d;y)$ are the generalized hypergeometric functions.

To compute the absorption length, it is necessary to consider sufficiently large radial distances in the wavefunction, i.e., $\left[u_{\omega m}(x)\right]_{x\rightarrow \infty}$. This can be done considering $z\rightarrow - \infty$ in Eq.~\eqref{eq:heunlf}, what leads to \cite{abramo}
\beqn
{\rm F}\left(1\pm\lambda, \pm \lambda; 1\mp \frac{imC}{r_{\rm h}};1\right)= \mp \frac{\Gamma\left(1\mp\frac{imC}{r_{\rm h}}\right)\Gamma\left(\mp m\right)}{\frac{m}{2}\left(1+\frac{iC}{r_{\rm h}}\right)\Gamma\left[\mp\frac{m}{2}\left(1+\frac{iC}{r_{\rm h}}\right)\right]^2},
\label{eq:hyperlf}
\eeqn
with $\Gamma(x)$ being the gamma function.

In the low-frequency regime, the absorption length is dominated by the $m=0$ mode. Thus, to find the absorption length in the low-frequency limit, we can consider only the mode $m=0$, and we can use that
\beq
\left\{\frac{\Gamma\left(1\mp\frac{imC}{r_{\rm h}}\right)\Gamma\left(\mp m\right)}{\frac{m}{2}\left(1+\frac{iC}{r_{\rm h}}\right)\Gamma\left[\mp\frac{m}{2}\left(1+\frac{iC}{r_{\rm h}}\right)\right]^2}\right\}_{m=0}=1.
\label{eq:gammalf}
\eeq

From Eqs.~\eqref{eq:wavefunction},~\eqref{eq:heunlf},~\eqref{eq:hyperlf} and~\eqref{eq:gammalf}, we find
\beq
u_{\omega m}\left(r\right)\propto r_h^2\sqrt{r},
\eeq
in agreement with the result obtained in Ref. \refcite{Oliveira:2010zzb}.

\subsection{Comparison with numerical results}
The absorption length of the draining bathtub was investigated numerically in Ref. \refcite{Oliveira:2010zzb}, using the partial waves method. Here we revisit this investigation, as a consistency check of our analytical solution. Following the procedure in Ref. \refcite{Oliveira:2010zzb}, we solve Eq. (\ref{eq:radeq}) numerically from $r=r_h+\delta r$, with $\delta r \ll 1$, to $r\gg r_h$, using Eq. (\ref{eq:hor}) and its derivative as boundary conditions. To find the absorption cross section, we use Eq. (\ref{eq:inf}) and its derivative to find the reflection coefficient, which we then substitute in Eq. (\ref{eq:mab}).

\begin{figure}
\centering
\includegraphics[width=0.7\columnwidth]{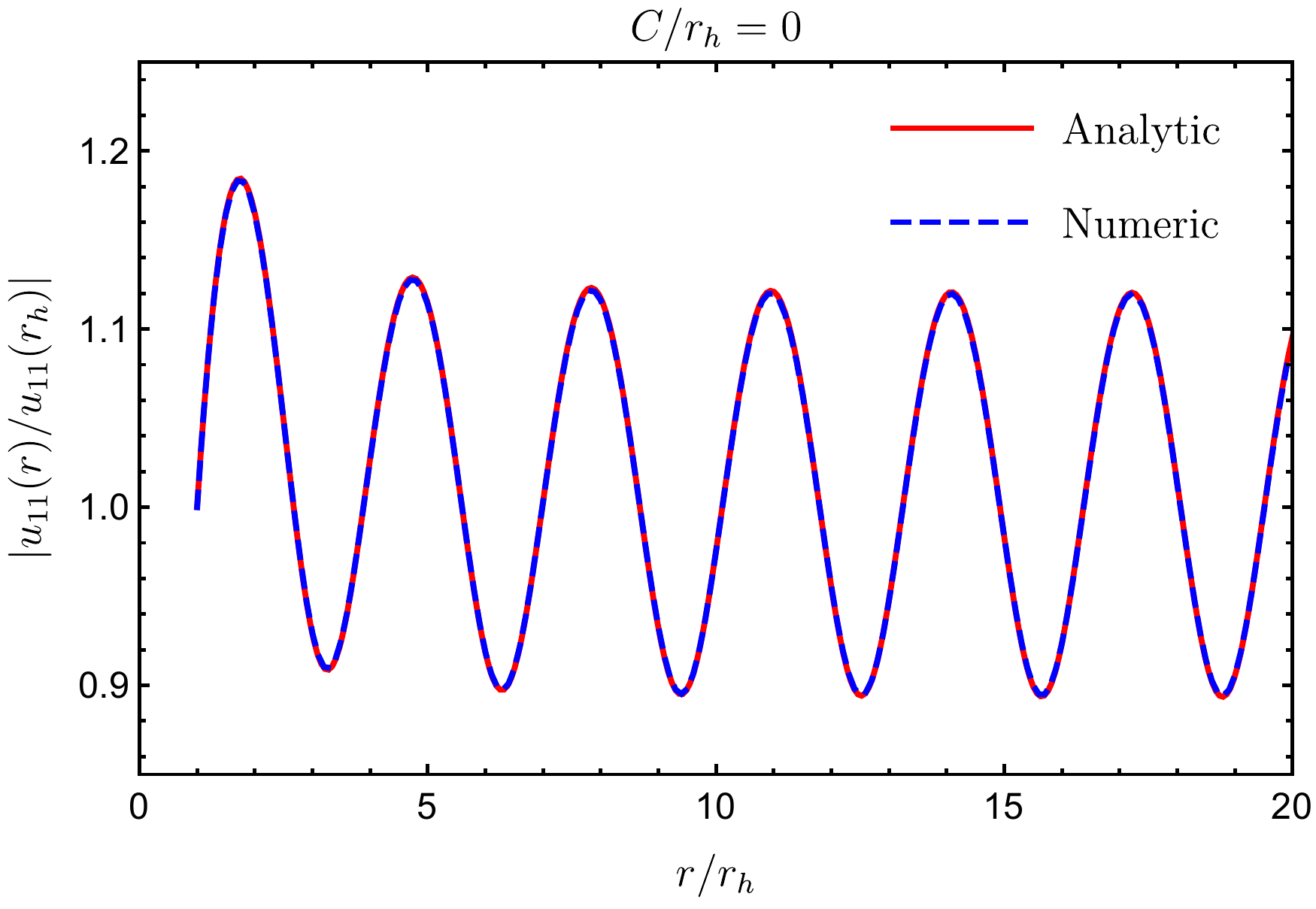}\vspace{0.3cm}\\
\includegraphics[width=0.7\columnwidth]{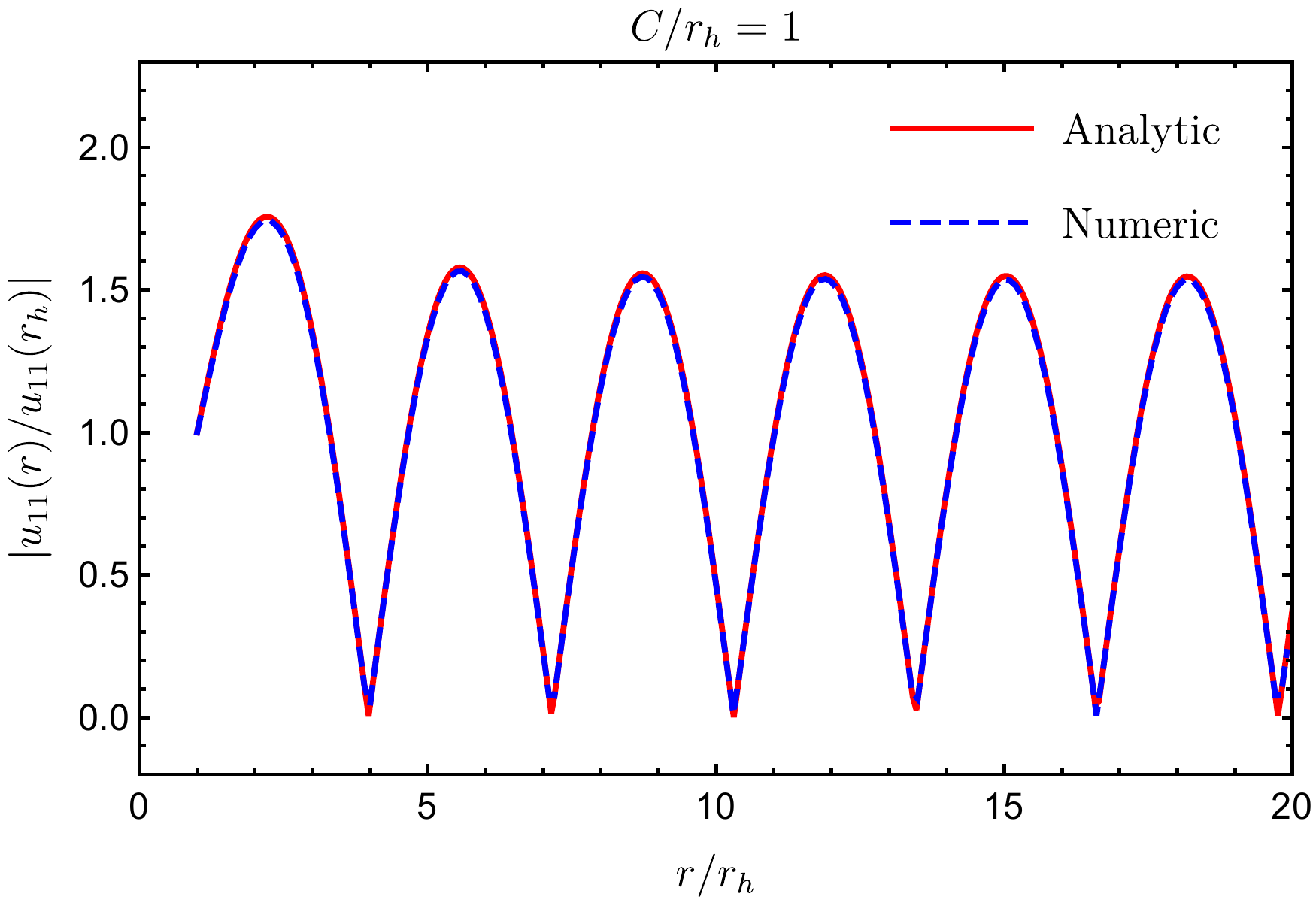}\vspace{0.3cm}\\
\includegraphics[width=0.7\columnwidth]{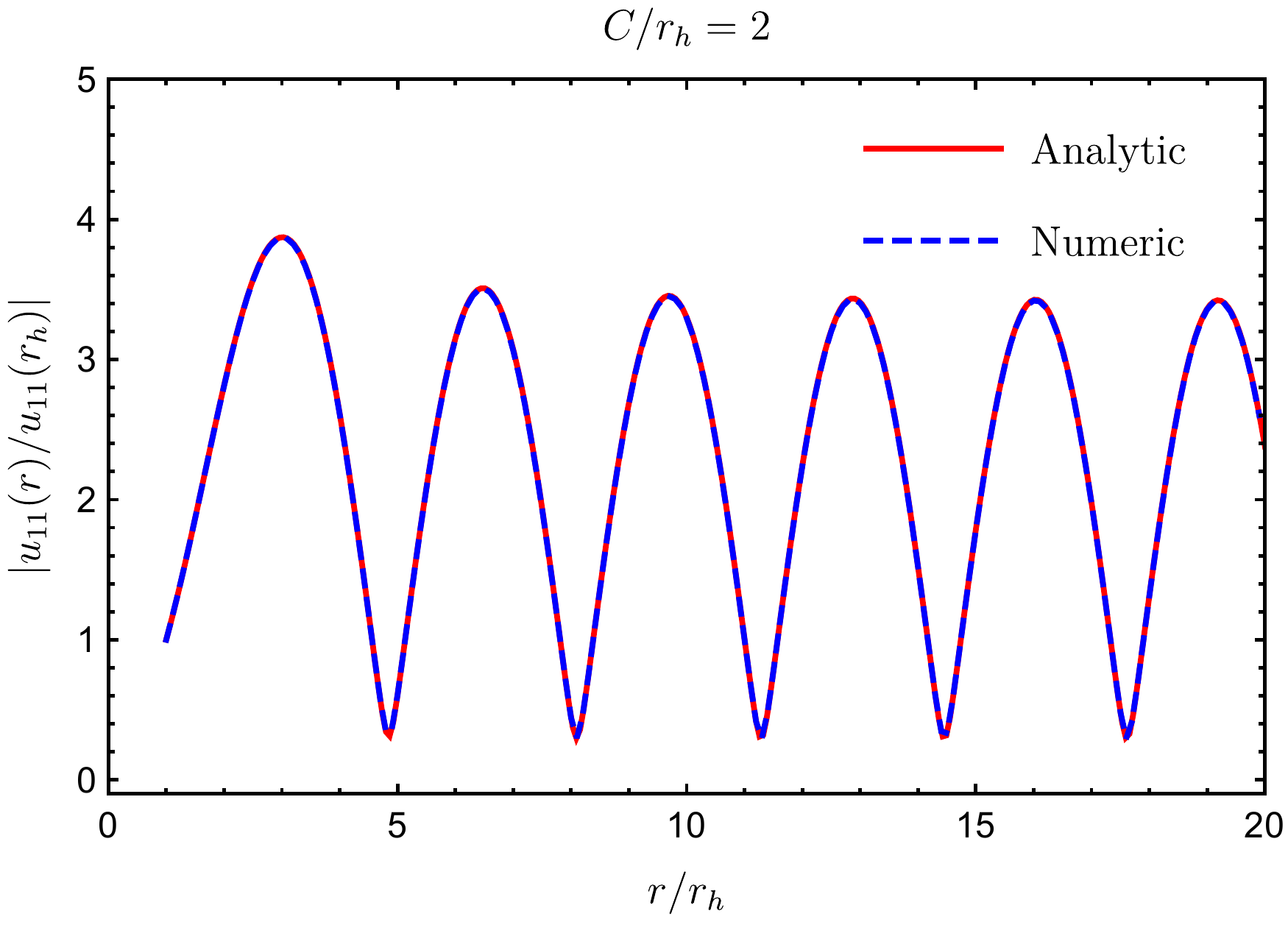}
\caption{Absolute value of the radial wavefunction, as a function of $r$, for $m=1$, $\omega=1$ and circulations $C=0$ (top), $C=r_h$ (middle) and $C=2r_h$ (bottom). The radial profile is typical of a propagating wave.}
\label{fig:wave}
\end{figure}

In Fig.~\ref{fig:wave} we compare analytical and numerical results of the absolute value of the radial part of the wavefunction, $\left|u_{\omega m}(r)\right|$, as a function of $r$, for a specific choice of the wave frequency and azimuthal number $m$, and different values of the circulation $C$. The analytic results were obtained using Eqs.~\eqref{eq:res}, \eqref{eq:ab} and \eqref{eq:wavefunction}. The constants $K_1$ and $K_2$ were fixed using the boundary condition exhibited in Eq. (\ref{eq:hor}), and we have chosen to omit their explicit forms, which are lengthy. The analytical and numerical results are in excellent agreement.

\begin{figure}
\centering
\includegraphics[width=0.7\columnwidth]{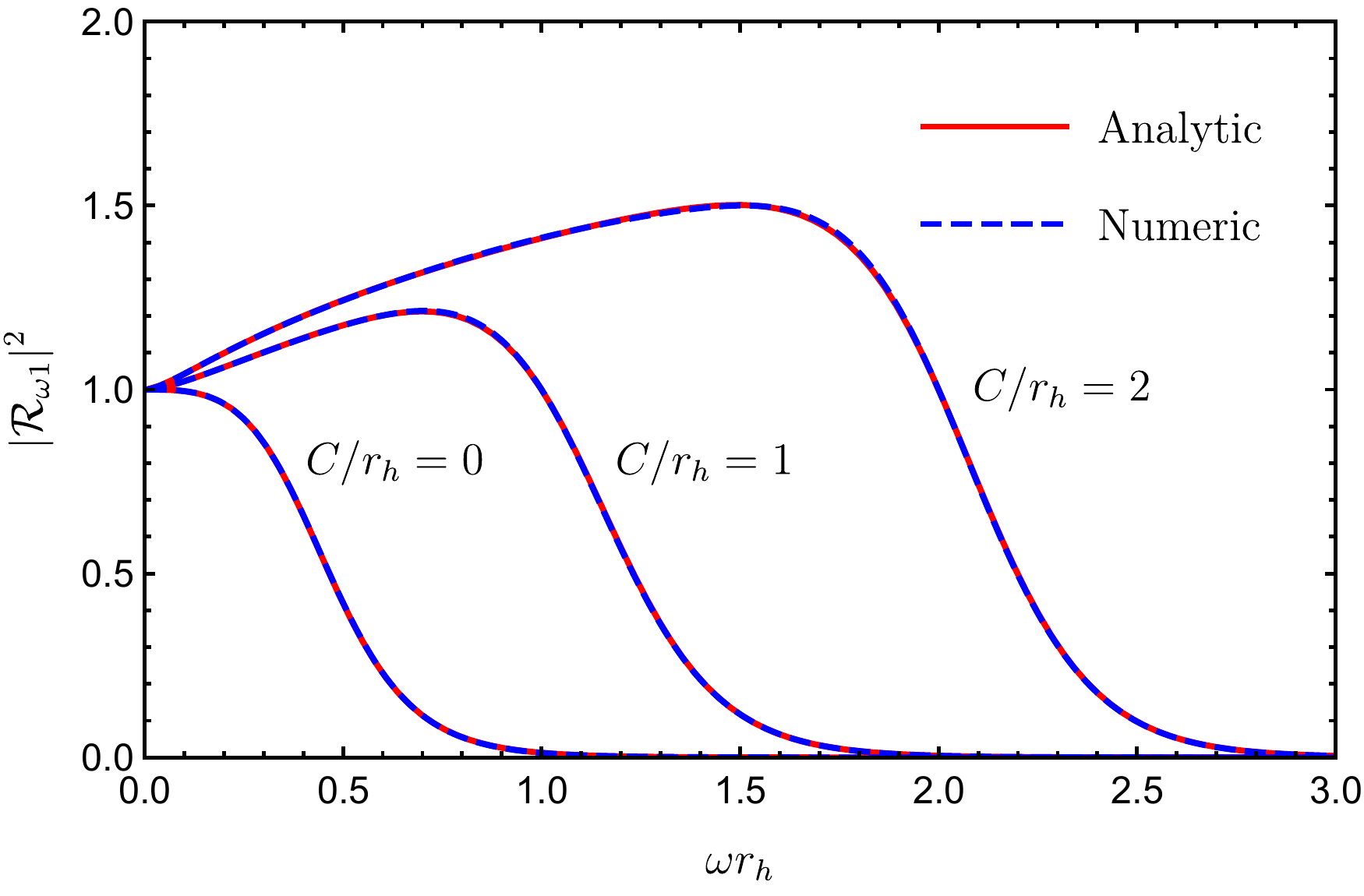}
\caption{Comparison of the reflection coefficient with $m=1$, obtained with the numerical and analytical solutions of Eq. (\ref{eq:res}),  for different values for the circulation. As the circulation increases, we can see the superradiance effect (which leads to a reflection coefficient greater than 1) increases.}
\label{fig:ref}
\end{figure}

\begin{figure}
\centering
\includegraphics[width=0.7\columnwidth]{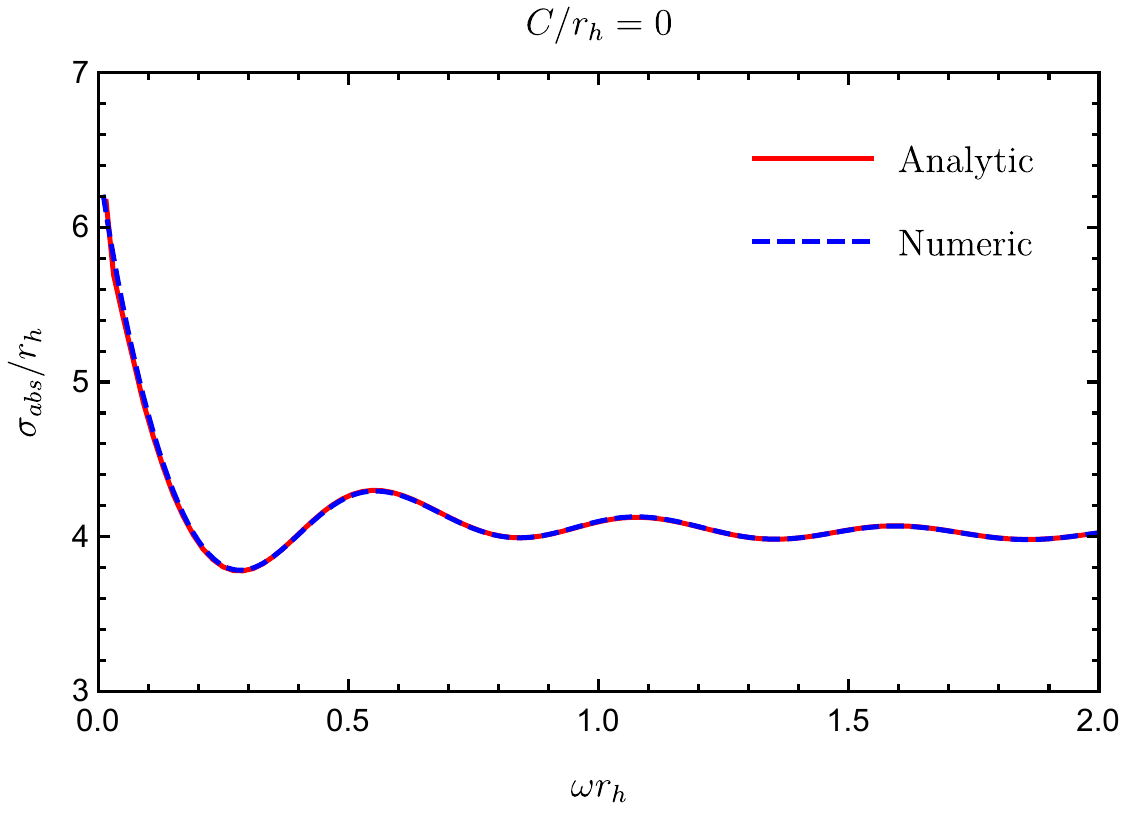}\vspace{0.3cm}\\
\includegraphics[width=0.7\columnwidth]{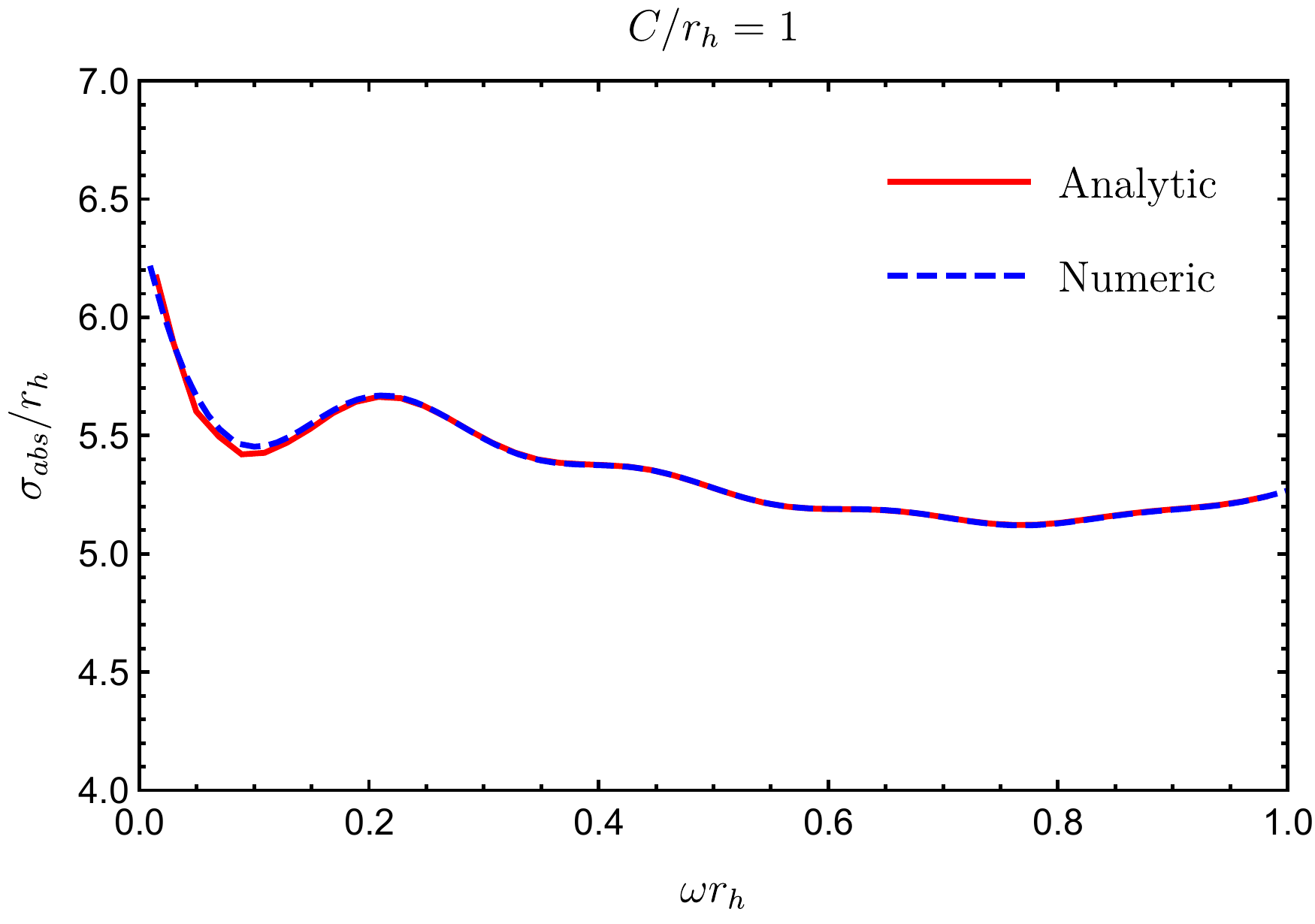}\vspace{0.3cm}\\
\includegraphics[width=0.7\columnwidth]{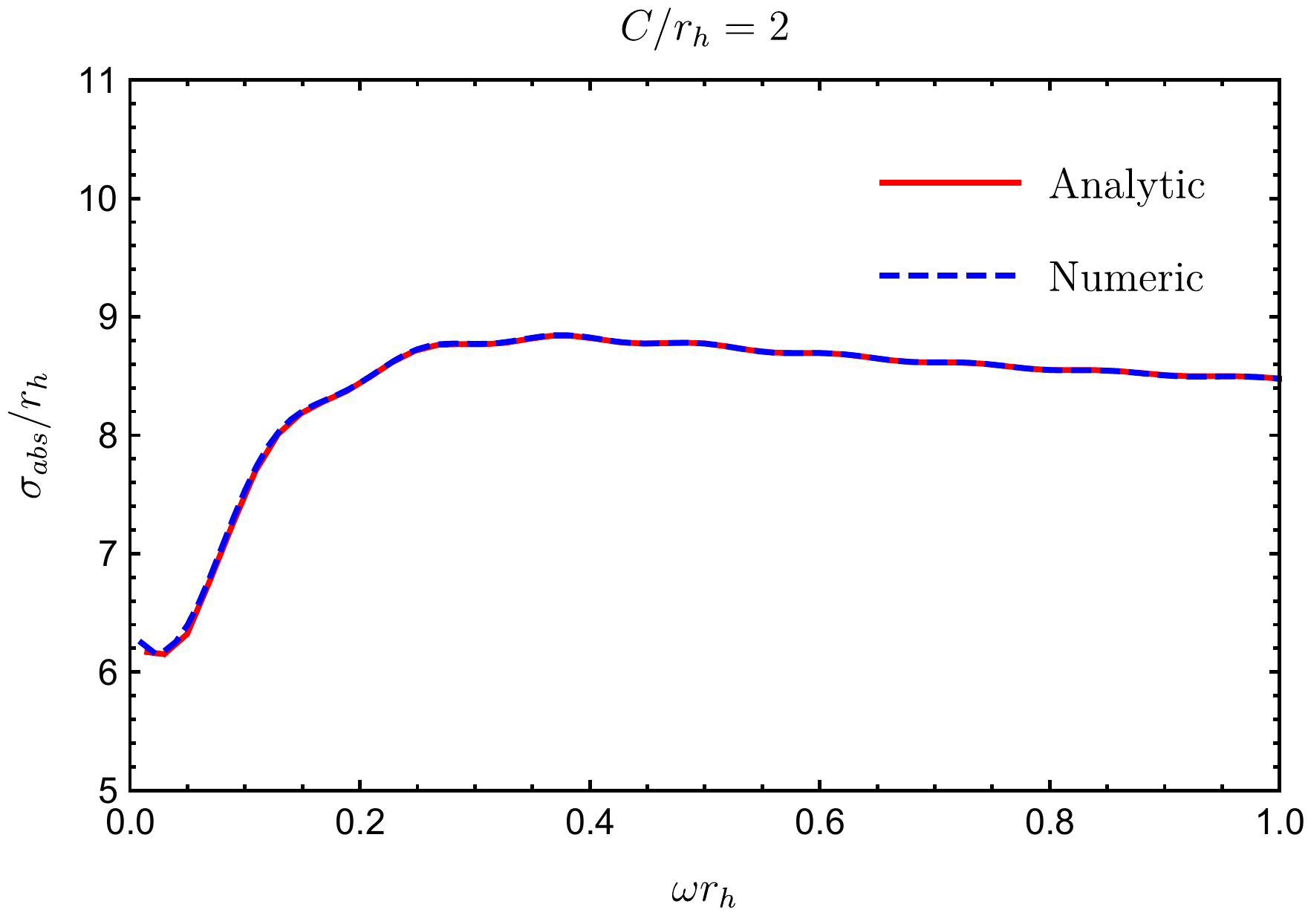}
\caption{Absorption length of the draining bathtub, $\sigma_{\rm abs}$, as a function of the frequency $\omega$. 
For this plot we have considered the summation of azimuthal numbers from $m=-9$ up to $m=+9$, and circulations $C=0$ (top), $C=r_h$ (middle)  and $C=2r_h$ (bottom). As the circulation increases, we have less regular oscillations, due to the frame-dragging effect.}
\label{fig:absm}
\end{figure}

In Fig.~\ref{fig:ref} we display the analytical and numerical results for the reflection coefficient, $\left|\mathcal{R}_{\omega m}\right|^2$, for circulation $C=0$, $r_h$ and $2r_h$. We see that for certain values of the circulation the reflection coefficient is greater than 1, what happens due to superresonance \cite{Basak:2002aw}. In Fig.~\ref{fig:absm} the results for the absorption length, as a function of the frequency $\omega$, are exhibited.
In both Figs.~\ref{fig:ref} and~\ref{fig:absm} analytical and numerical results agree remarkably well.

\section{Conclusion}
\label{sec:conclusion}

In this letter we have considered perturbations in a rotating BH analogue model, solving the corresponding equations analytically and numerically. We investigated the draining bathtub, that possess a draining and circulating fluid flow. When the speed of the fluid flow is high enough, perturbations in such a fluid feel an ergorregion and an event horizon. 

The absorption length of planar waves in this effective spacetime was investigated numerically in Ref.~\refcite{Oliveira:2010zzb}. However, it is possible to find analytical solutions by writing the radial solution in terms of Heun functions, as shown in Ref.~\refcite{Vieira:2014rva}. We have obtained analytically the absorption length in terms of the confluent Heun functions and performed consistency checks, comparing our results with the absorption length computed numerically, as well as with the low-frequency limit, obtaining great agreement for all cases. 

Our analytical solutions manifest the physical characteristics of the absorption by rotating BHs, i. e.: 
(i) the total absorption length presents an oscillating pattern, as the maximum contribution of each partial wave occurs for a different value of the frequency; 
(ii) as we increase the value of the circulation, the oscillating pattern becomes less regular (cf. Fig. \ref{fig:absm});
(iii) low-frequency modes can be amplified, due to superresonance (cf. Fig. \ref{fig:ref}); 
(iv) in the low-frequency limit the absorption length goes to the perimeter of the event horizon. 

Analytical investigations are always desirable, even for systems already studied numerically, since the analytical solutions may help us to have a better understanding of the physical problems. 
The Heun functions, in particular, appear as solutions in a variety of physical systems \cite{Ronveaux:1995,Hortacsu:2011rr}. They can be used, e. g., to investigate quasinormal modes of compact objects \cite{Fiziev:2011mm,Fiziev:2019ewy}, which are associated to the form of the gravitational waves coming from the merger of two compact objects, for instance. Further investigation in this direction can help in our understanding on the characteristics of compact objects and their interaction with fields.


\section*{Acknowledgments}

This research was financed in part by Coordena\c{c}\~ao de Aperfei\c{c}oamento de Pessoal de N\'ivel Superior (CAPES, Brazil) -- Finance Code 001, and by Conselho Nacional de Desenvolvimento Cient\'ifico e Tecnol\'ogico (CNPq, Brazil). This research has also received funding from the European Union's Horizon 2020 research and innovation programme under the H2020-MSCA-RISE-2017 Grant No. FunFiCO-777740. L. C. thanks Valdir B. Bezerra for useful discussions.

\end{document}